# Tree frog-inspired nanopillar arrays for enhancement of adhesion and friction

Zhekun Shi, Di Tan, Quan Liu, Fandong Meng, Bo Zhu, and Longjian Xue

View Online     Export Citation     CrossMark

# Tree frog-inspired nanopillar arrays for enhancement of adhesion and friction

Zhekun Shi, 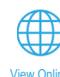 Di Tan, Quan Liu, Fandong Meng, Bo Zhu, and Longjian Xue[a])

### AFFILIATIONS

School of Power and Mechanical Engineering and The Institute of Technological Science Wuhan University,
South Donghu Road 8, Wuhan, Hubei 430072, China

**Note:** This paper is part of the Biointerphases Special Topic Collection on Biomimetics of Biointerfaces.
[a)]**Electronic mail:** xuelongjian@whu.edu.cn

### ABSTRACT

Bioinspired structure adhesives have received increasing interest for many applications, such as climbing robots and medical devices. Inspired by the closely packed keratin nanopillars on the toe pads of tree frogs, tightly arranged polycaprolactone nanorod arrays are prepared by mold process and chemical modification. Nanorod arrays show enhanced adhesion and friction on both smooth and rough surfaces compared to the arrays with hexagonal micropillars. The bonding of nanorods results in a larger stiffness of the nanorod surface, contributing mainly to friction rather than adhesion. The results suggest the function of closely packed keratin nanopillars on the toe pad of tree frogs and offer a guiding principle for the designing of new structured adhesives with strong attaching abilities.

Published under license by AVS. https://doi.org/10.1116/6.0000747

## I. INTRODUCTION

Micro- and nanostructured adhesives inspired by the pillar-like structures on the toe of animals like geckos and frogs have drawn great attention in the past two decades.[1–11] The geometry of the pillar array allows conformal contacts to rough surfaces, generating strong adhesions.[12–14] Compared to the great attention on geckos, much less attention has been paid to the tree frogs which are able to strongly attach and smoothly climb on rough/smooth surfaces under dry/wet conditions.[5,8,15] Patterned surfaces on the toe pads are demonstrated to act a critical role in their performance.[8,16–18] The toe pad surfaces are mainly composed of polygonal epithelial cells (∼10-$\mu$m side length) separated by ∼1-$\mu$m wide microchannels.[18,19] The surface of epithelial cells in *Staurois parvus* is found to be covered by densely packed keratin nanopillars (∼300–400 nm in diameter) with a nanoscale concave on the tip.[19] However, the understanding of the contribution of these closely packed nanopillars to the adhesion and friction performance is rather limited.[8,20,21]

Inspired by the structured surfaces on tree frog toe pads, micro- and nanostructured surfaces have been composed and investigated.[11,15,20,22–28] Compared to the flat surface, hexagonal micropillar arrays presented a higher friction force on wet surfaces due to the formation of direct contacts by draining the liquid out of the contact interface and the effect of contact splitting.[22,23,29] For the hexagonal micropillars, the sizes of the micropillar and the channel have a strong influence on the adhesion performances. The effective contacts will be reduced by increasing the channel width, resulting in a smaller adhesive force.[27] Meanwhile, the shape of micropillars and the corresponding sliding direction (corner-sliding or edge-sliding) also have a strong influence on the efficiency of liquid drainage that elongated hexagonal[22,24] and arch-shaped[28] micropillars possessed larger friction forces than hexagonal micropillars on wet surfaces. Recently, the incorporation of nanocavities on top of micropillars improved the boundary friction over 3.5 times as compared to a smooth surface.[25]

On the other hand, closed packed, vertically aligned keratin nanofibrils are widely found on the toe pad of the frog species bearing micropatterned adhesive pads.[2,10,19,21] Inspired by these nanofibrils, we successfully constructed a composite micropillar array composed of polydimethylsiloxane (PDMS) micropillars embedded with vertically aligned polystyrene (PS) nanopillars.[20] The composite micropillar array showed enhanced adhesion and friction than the pure PDMS array, which is originated from the homogenized but discretely distributed stress at the contact interface. Mixing the PS nanoparticles into PDMS micropillars, followed by the removal of PS nanoparticles on the surface, composite micropillars with nanopits on the micropillar top were successfully fabricated.[26] The composite micropillars showed ~36.5 times wet adhesion of a tree frog's toe pad, where the micro- and nano-sized liquid bridges at the interface play a significant role.[26] These previous works suggest the important role of these aligned keratin nanofibrils in the adhering ability of tree frogs in dry and wet environments. However, the understanding of the functions of these closely packed keratin nanopillars remains far from satisfactory.[5,8,15]

Here, we design a series of polycaprolactone (PCL) nanorod arrays (termed NRA) in order to mimic the closely packed keratin nanofibrils on tree frog toe pads and to understand their possible influence on adhesion and friction. The NRA was shaped into a hexagonal pattern by embossing and/or modified with dopamine. The dimensions of the hexagonal pattern are close to the tree frog toe pads. Enhanced adhesion and friction performances of series of NRA were measured on both smooth and rough surfaces, as compared with the hexagonal micropillars. The results suggest the important role of the tight arrangement of keratin nanopillars on the toe pads of tree frogs in the realization of strong attachment, especially in friction performance. The study here offers a guiding principle for the design of bioinspired fibrillar adhesives for rough and smooth surfaces.

## II. EXPERIMENT

### A. Materials and equipment

Polycaprolactone (PCL, average $M_w$ is about 45 kDa), $SiO_2$ particles (250 nm in diameter), and dopamine hydrochloride were obtained from Sigma–Aldrich. PCL has a modulus of ~220 MPa and a melting point of ~60 °C, which allows an easy fabrication.[30] On the other hand, if a soft material like polydimethylsiloxane (PDMS) is used, the pillars in the nanoscale will collapse. If a hard material like polystyrene (PS) is used, then the design will not conform to the soft tree frog toe pad. The surface microstructures were characterized using a scanning electron microscope (SIGMA, Zeiss AG, Germany) and a white-light interferometer (NewView 9000, ZYGO Corp., USA). The topography of PCL nanopillar arrays was sputter-coated with a gold layer before SEM imaging. Plasma treatment of samples was carried out with a Plasma Activate Studio 10 USB chamber (Plasma Technology GmbH, Rottenburg, Germany). The water contact angles were measured on a KRÜSS DSA10-MK2 (KRÜSS GmbH, Germany) drop shape analysis system with $3 \mu l$ of de-ionized water as the probe fluid. The average value of the water contact angle was obtained by measuring the same sample at three different locations.

### B. Fabrication of adhesives

The tree frog toe pad-inspired micro- and nanopillars adhesives were fabricated by soft lithography and chemical modification, as shown schematically in Fig. 1. Self-ordered anodic aluminum oxide (AAO) templates were fabricated by two-step anodization in phosphoric acid as reported previously.[31] The PCL pellets were placed on AAO and heated to 65 °C for 20 min under vacuum. The aluminum layer in the AAO template was dissolved by immersion in a solution of 100 ml of 37% HCl and 3.4 g of $CuCl_2 \cdot 2H_2O$ $CuCl_2 \cdot 2H_2O$ in 100 ml of de-ionized water at 0 °C. After the removal of the aluminum layer, the alumina layer was etched away in 1 M aqueous NaOH solution at room temperature for 1 h, and the NaOH solution was replaced with a fresh one for another 1 h. The PCL nanorod array (termed NRA) was obtained after drying in a freeze-dryer [Fig. 1(a)]. By using a nickel shim with a negative copy of hexagonal micropillar pattern ($20 \mu m$ in diameter, $5 \mu m$ in depth, and $3 \mu m$ in the gap between the holes) [Fig. 1(b)], the NRA sample was further shaped into a hexagonally micropatterned nanorod array under proper pressure, termed HexNRA [Fig. 1(c)]. The molding process crushed the NRA selectively below the walls of the Ni shim but not within the holes. In this method, a hexagonal microchannel pattern was superimposed onto NRA, where the height of the resulted micropillar is consistent with NRA ($5 \mu m$). Then, NRA was treated with oxygen plasma and subsequently immersed in a solution of dopamine hydrochloride dispersed in the alkaline aqueous solution for 30 min. After freeze-drying, the dopamine modified NRA (termed dopaNRA) was achieved [Fig. 1(d)]. Dopamine, a natural catecholamine, can be polymerized into polydopamine (PDA) in base condition. PDA can adhere to almost all the surfaces, such as ceramics, noble metals, metal oxides, and polymers (including Teflon).[32] Hence, dopamine was used to bond neighboring PCL nanorods. The modification process is simple, convenient, and efficient. Likewise, the nanorod array with hexagonal micropattern with dopamine modification, termed as dopaHexNRA, was also fabricated in the same way [Figs. 1(d)–1(f)]. The nickel mold was also employed as the template to prepare the hexagonal micropillar array [Figs. 1(g) and 1(h)] by annealing at 65 °C for 20 min under vacuum. After cooling to room temperature, the sample was then ready to peel off from the nickel mold and is denoted as Hex.

### C. Surface modification of the probe

A glass sphere was coated by $SiO_2$ particles to mimic rough surfaces (Fig. 2). The glass sphere was treated with oxygen plasma with 100 W, 0.1 mbar for 30 s. A droplet solution containing $SiO_2$ nanoparticles (250 nm in diameter) was dropped on the surface of the glass sphere, which was subsequently annealed in an oven at 650 °C for 10 min to sinter the $SiO_2$ nanoparticles and to enhance adhesion of these $SiO_2$ nanoparticles to the glass sphere.

### D. Adhesion and friction tests

Adhesion and friction tests were all conducted in ambient conditions on a custom-made device,[24,33,34] which employs a spherical glass with a diameter of 5 mm or the glass probe decorated with $SiO_2$ nanoparticles as the probe to simulate smooth and rough surfaces, respectively. The minimum resolvable force increment, maximum force, and the temporal resolution of the sensor are 0.1%, 100 mN, and $1.5 \mu s$, respectively. The probe is connected to the upper force sensor which controls the loading forces in adhesion and friction tests. The sample was mounted on the lower sensor, which records the lateral friction force. The probe was cleaned with acetone before testing. In adhesion tests, the probe approached the sample at a constant speed of $20 \mu m/s$ until a predefined loading force (1–5 mN) was reached and then retracted in the opposite direction to detach from the surface at the same speed. The adhesion force corresponds to the maximum force required for detachment from the samples. In friction tests, the probe was brought into contact with the sample surface and a normal force was applied and kept constant during the lateral shearing. The sample was moved at a velocity of $100\,\mu m/s$ over a distance of $500\,\mu m$, forward and backward, while the forces were simultaneously recorded. Both adhesion and friction measurements were repeated three times on three samples. For data collection and processing, a custom-made program in LABVIEW software was used.

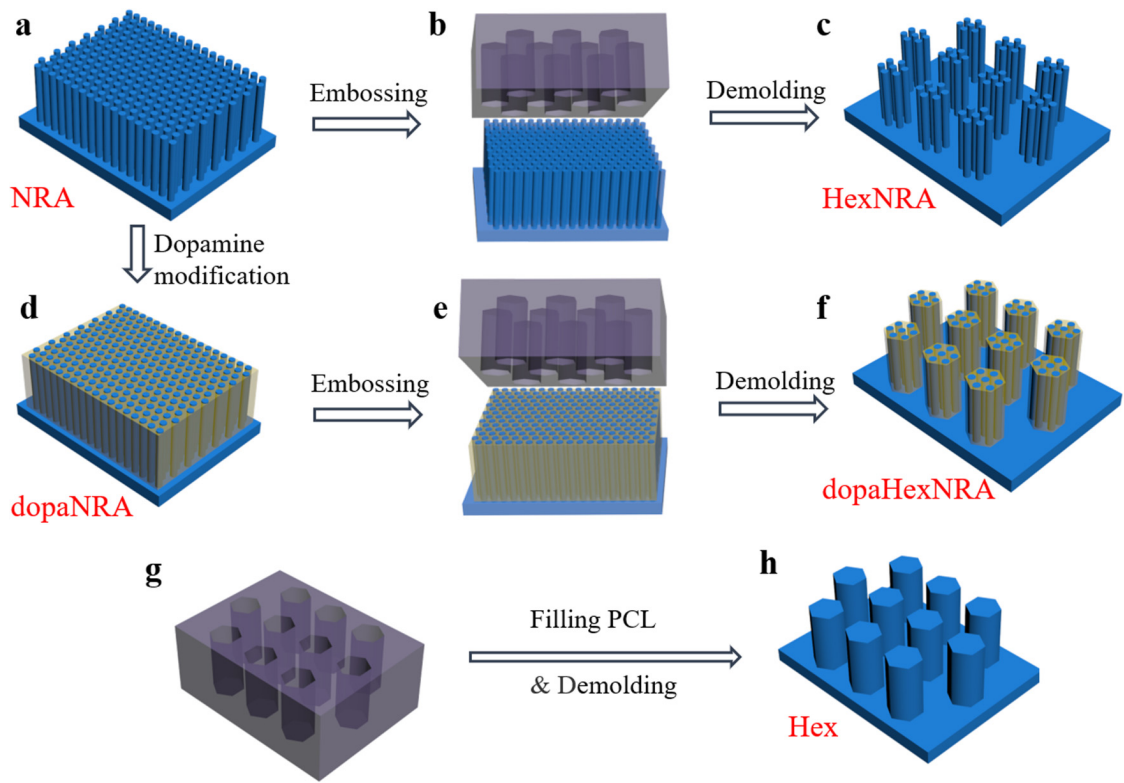

**FIG. 1.** Schematic illustration of the fabrication process of tree frog-inspired nanorod arrays. (a) PCL nanorod array (NRA, blue) replicated from the AAO template. (b) Embossing with nickel mold insert (purple) on NRA array at high pressure. (c) Hexagonal micropatterned PCL nanorod array formed after embossing, termed HexNRA. (d) PCL nanorod array modified with dopamine, termed dopaNRA. (e) Embossing with nickel mold insert (purple) on dopaNRA array at high pressure. (f) Dopamine-modification hexagonal micropatterned PCL nanorod array formed after embossing, termed dopaHexNRA. (g) Nickel template with hexagonal holes. (h) Hexagonal PCL micropillar pattern replicated from the Ni template, termed Hex.

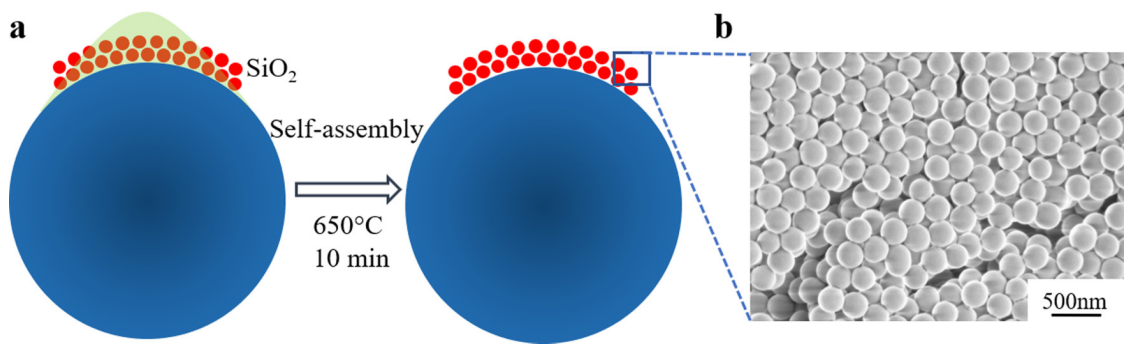

**FIG. 2.** (a) Schematic illustration of the fabrication of rough probe by coating SiO$_2$ nanoparticles on to a glass ball. (b) SEM image of SiO$_2$ assembled on the spherical probe.

## III. RESULTS AND DISCUSSION

### A. Construction of tree frog-inspired toe pads

Five different kinds of tree frog toe pad-inspired adhesives were fabricated by templating and chemical modification. PCL NRA with a period of 500 nm, diameters of 330 nm, and height of 5 $\mu$m was successfully replicated from the AAO template [Figs. 3(a) and 3(b)]. The dimension of the nanorods matches very well with the keratin nanopillars found in tree frog toe pads.[19] The dopaNRA [Fig. 3(c)] shows a water contact angle (ca. 60.7°), indicating a successful coating of dopamine as the water contact angle of NRA was ca.66.7°.[35] The dopaNRA was tightly connected to each other [red arrow in Fig. 3(d)]. It, thus, offers us the opportunity to mimic the state of the keratin fibers in the tree frog toe pad which are always having some connections to the keratin fibers around. If connected, then the bundles of fibers, thus, possess a larger elastic modulus compared with the scattered nanorods in NRA. The concave [red circle in Fig. 3(d)] on top of the nanorods is also a mimicking of the nanoconcaves on the tree frog toe pad. It has been proposed that they are helpful to build up nanoscale liquid bridges at the contact interface and may generate a suction effect under a relatively large loading force.[26] The dopaHexNRA shows a hexagonal micro-patterned nanorod array after molding [Fig. 3(e)]. The bottoms of the channel between the micropillars have a flat surface, indicating the pressure was large enough to crush the nanopillars below the walls of the Ni shim; on the other hand, the standing configuration of remaining dopamine-modification PCL nanorods suggests that the pressure was not too large to destroy the nanorods within the holes. Note that the top surface of the micropillar is smooth in Hex [Fig. 3(f)] compared with HexNRA and dopaHexNRA.

### B. Performance against the smooth surface

The adhesion and friction performances of the tree frog-inspired structured surfaces were examined using a spherical glass probe with a roughness of 0.3 nm [Fig. 4(a)]. A zigzag rising curve [Fig. 4(b)] appeared during the loading process. This can be attributed to the continuous deformation of PCL nanorods as the probe gradually compresses into the array. As PCL has a relatively small elastic modulus of 220 MPa,[50] the PCL nanorods at the center of the contact area start to deform once they get into contact with the spherical probe. The further pressing brings more nanorods into contact. On the other hand, due to off-centered compression (probe against the PCL nanopillar) and the tilting/bending of PCL nanopillars, the large deformation of the contacted nanorod may cause itself to slide away from the contacted position, resulting in the temporary dropping of the loading force. Macroscopically, a zigzag feature of loading was formed. After loading, it was followed by an immediate retraction of the probe at the same speed. The adhesion force ($F_{ad}$) corresponds to the value of force that separates the probe from the measuring sample.

The adhesion performances at different loading forces (1–5 mN) were evaluated for all arrays. It should be stated that no liquid was added to the contact interface in all tests here. Previous work has demonstrated that the tree frog-inspired toe pads can drain liquid out from the contact interface and form solid–solid direct contact on a wet surface where van der Waals forces may contribute mainly to the adhesion forces.[24] Therefore, we believe the difference in surface energy originated from the modification of dopa (dopaHexNRA and dopaNRA) has an ignorable influence on the adhesion performance. The current work, therefore, focuses on the understanding of the contribution of states of nanorods to the adhesion and friction on dry surfaces. A larger $F_L$ results in a larger contact area (more contacting nanorods) and consequently leads to an improvement of adhesion force.[3,33,36] The correlation coefficients (R) of dopaHexNRA, HexNRA, dopaNRA, NRA, and Hex are 0.608, 0.979, 0.958, 0.968, and 0.969, respectively [Fig. 4(c)], confirming the strong influence of $F_L$. $F_{ad}$ of NRA increased from $0.07 \pm 0.03$ to $0.16 \pm 0.05$ mN when $F_L$ increased from 1 to 5 mN, showing better adhesion performance than the other arrays. Slightly worse adhesion performance was found on HexNRA, as less contact points formed between the nanorods and the probe due to the existence of channels. The surface with sparse/less nanopillars has an even smaller effective elastic modulus. Therefore, under the same loading force, the surface with less nanopillars (considering the nanopillars are homogeneous on the surface) will have a larger indentation depth, which means a larger contact area. However, fewer nanopillars will be involved within the contact area due to the existence of microchannels. Therefore, the number of contact points formed is reduced, resulting in a smaller van der Waals force. At $F_L = 5$ mN, $F_{ad}$ of NRA is 6 times and 2.6 times of dopaHexNRA and Hex, respectively. The modification of dopamine bundles the nanorods together that the dopaHexNRA has a larger modulus than the scattered nanorods (NRA). Previous works have demonstrated that the material with a smaller elastic modulus is beneficial for the formation of reliable contacts and, thus, can generate stronger adhesion.[9,37] Consequently, the adhesion performance of dopaNRA is not as good as NRA. Therefore, it is reasonable that dopaHexNRA and Hex possess even worse adhesion. On the other hand, it is interesting to note, $F_{ad}$ of dopaHexNRA shows much less dependence on $F_L$. It infers that the connection among keratin nanofibrils may be beneficial for the adhesion performances of tree frogs which may allow the tree frogs to form reliable contacts during the locomotion with less effort.

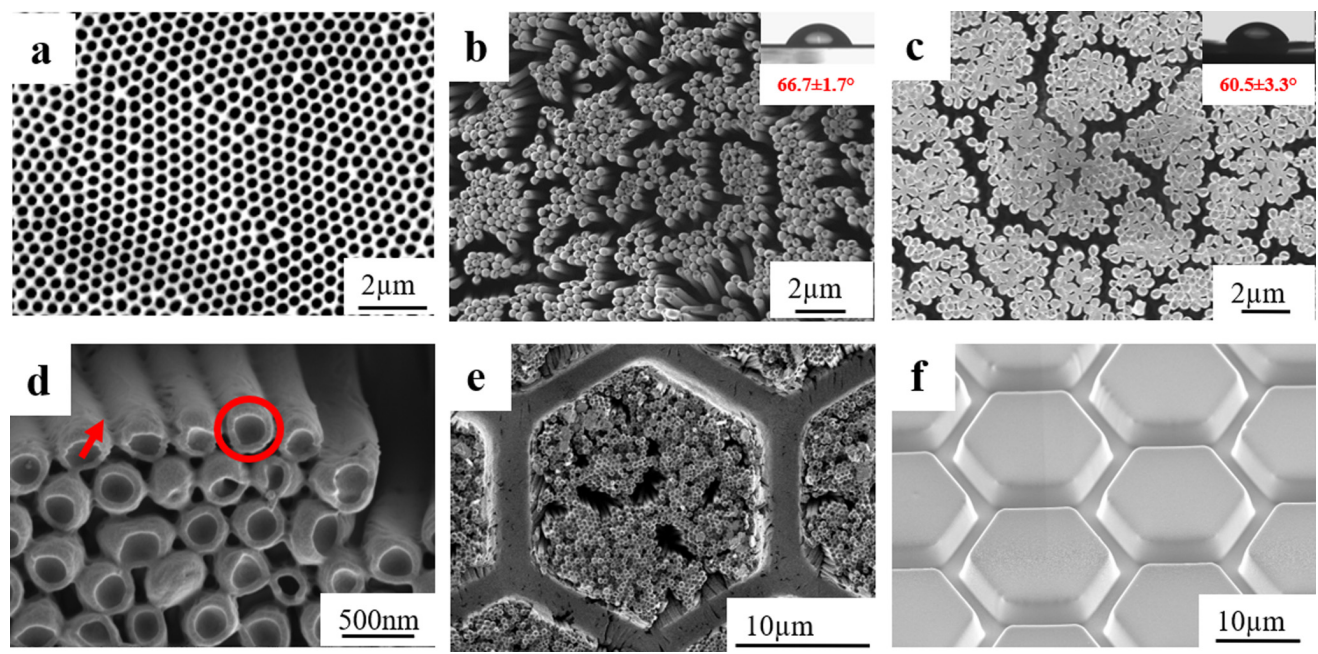

FIG. 3. SEM image of (a) the AAO template with a pore diameter of 330 nm, (b) NRA, (c) dopaNRA, (d) magnified image of dopaNRA showing the bonding with dopamine between NRA (red arrow) (e) dopaHexNRA and (f) Hex. The insets in (b) and (c) are the water contact angle of the corresponding array. The red circle in (d) indicates the concave tip of the PCL nanorod.

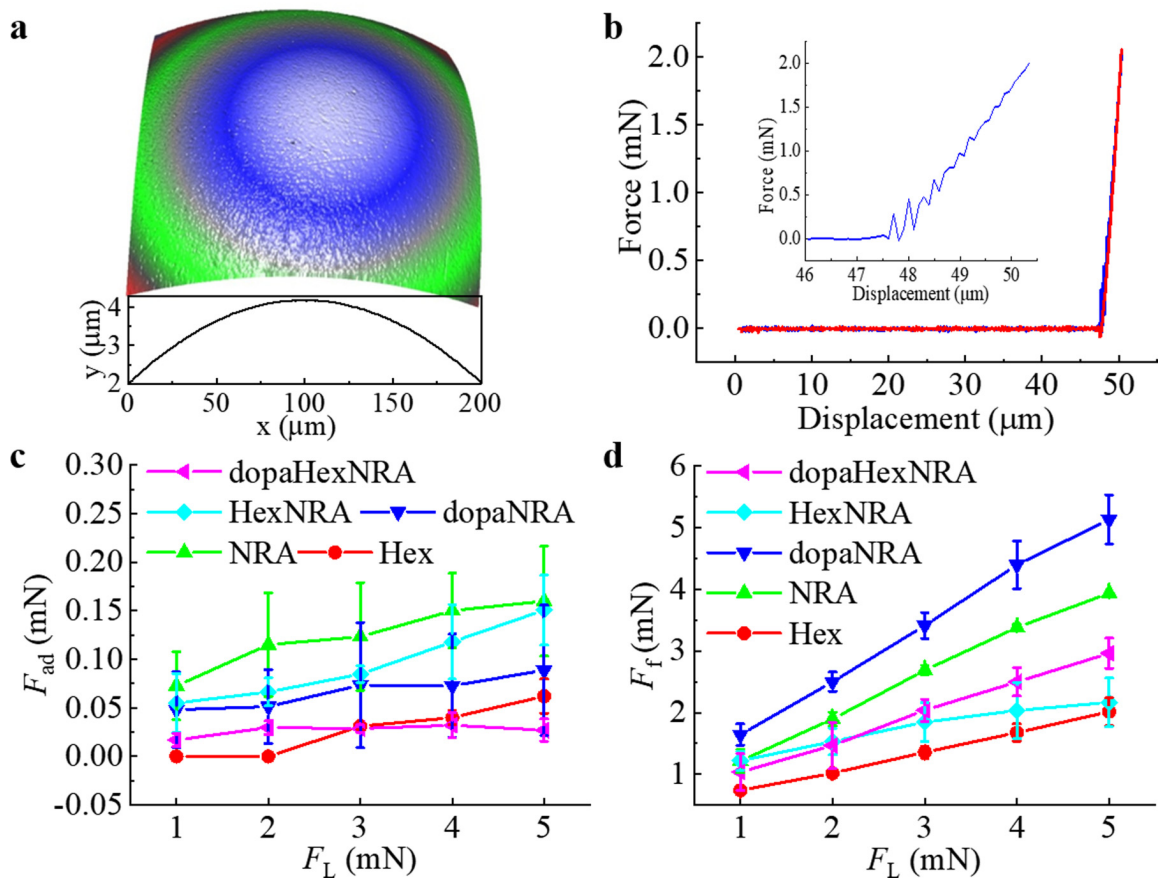

**FIG. 4.** (a) 3D morphology of the spherical probe with a typical profile. (b) Representative force–displacement curve of adhesion test with the probe shown in (a). The inset is the zoomed-in loading curve. Dependence of (c) adhesion force ($F_{ad}$) and (d) friction force ($F_f$) on the loading force ($F_L$) against the smooth probe. The values are averaged based on nine measurements on three samples. Error bars are the standard deviation.

For instance, $F_f$ of dopaNRA increased from $1.64 \pm 0.17$ to $5.13 \pm 0.39$ mN when $F_L$ increased from 1 to 5 mN, showing a friction coefficient of ~0.87, much higher than that of NRA (0.68), dopaHexNRA (0.48), HexNRA (0.24), and Hex (0.31). The slope of the linearly fitted line, where the coefficient of determination ($R^2$) of dopaHexNRA, HexNRA, dopaNRA, NRA, and Hex are 0.998, 0.970, 0.998, 0.997, and 0.999, respectively [Fig. 4(d)], indicates that $F_f$ is more sensitive than $F_{ad}$ to $F_L$ on a smooth surface. At $F_L = 5$ mN, $F_f$ of dopaNRA is 2.5 times of Hex ($2.01 \pm 0.22$ mN). Note that dopaNRA showed the better friction performance than NRA which showed the best adhesion performance. For friction, in addition to the surface contact area, the bending stiffness of the pillars has a significant influence on the friction performance. The bonding of the nanopillars by dopamine resulted in an increase in stiffness (modulus), which endows the pillars' larger resistance to bending when subjected to a lateral shearing. Thus, dopaNRA with larger stiffness and contact area showed a larger friction force than NRA. $F_f$ of dopaHexNRA was lower than dopaNRA due to the reduced contact area and was higher than HexNRA due to the bonding effect of the pillars. Hex possessed the worst fiction performance for its microscale pillars which provided less effective contact area. It is worth mentioning that the structural modulus contributes differently to $F_{ad}$ and $F_f$. A smaller modulus is better to form good contact and thus resulted in a higher adhesion but it is not beneficial for strong friction.[10] For instance, NRA showed stronger $F_{ad}$ but weaker $F_f$ than dopaNRA. Meanwhile, the nanostructured surfaces, like HexNRA, showed stronger $F_{ad}$ and $F_f$ than Hex. Together with the previous reports,[20,26] it strongly suggests the importance of the nanofibrils to the adhesion and friction of tree frogs.

## C. Performance against the rough surface

The performance of tree frog-inspired toe pads on the rough surface was also studied, as natural surfaces are always rough. A glass probe coated with silica nanoparticles (roughness 273 nm) was used to mimic the rough surface [Fig. 5(a)]. It should be noted that the size of the nanoparticles is close to the diameter of nanopillars. Once the nanorod array (uniform height) gets into contact with the rough surface, a single nanorod may touch the nanoparticle or locate in the gap between nanoparticles. The increase in loading force may cause the nanorod to slip away from the contacting nanoparticle into the gap between nanoparticles and/or bending of nanorods. Therefore, similar fluctuation but with larger amplitude in $F_L$ was found as compared to that against the smooth surface during the loading process [Fig. 5(b)].

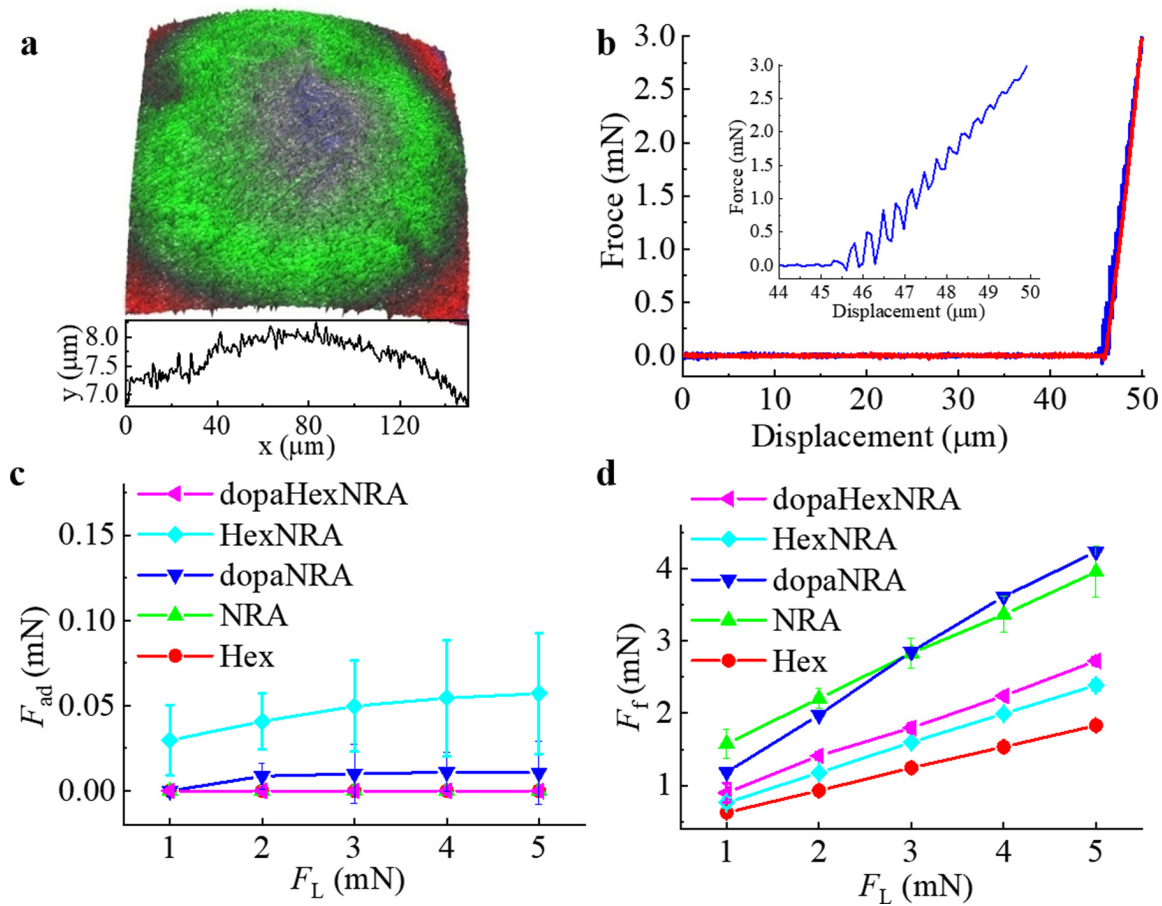

**FIG. 5.** (a) 3D morphology of the rough probe with a typical profile. (b) Representative force–displacement curve of adhesion test with the probe shown in (a). The inset is the zoomed-in loading curve. Dependence of (c) adhesion force ($F_{ad}$) and (d) friction force ($F_f$) on the loading force ($F_L$) against the rough probe. The values are averaged based on nine measurements on three samples. Error bars are the standard deviation.

Adhesions of all arrays significantly decreased on the rough surface due to the decline of effective contact area compared with the smooth surface [Fig. 5(c)]. Almost no adhesion was detected on all structures, except HexNRA. This is could be due to the mismatching between the period of the nanorods [500 nm, Fig. 3(d)] and $SiO_2$ [250 nm, Fig. 2(b)]. On the other hand, HexNRA has the best adhesion performance that $F_{ad}$ increased from $0.029 \pm 0.02$ to $0.057 \pm 0.03$ mN when $F_L$ increased from 1 to 5 mN. It suggests that microscale hexagonal patterns with internal nanorods array may have a much larger chance to form effective contact with the rough surfaces. The gap between the hexagonal patterns allows nanopillars at the perimeter of the hexagonal pattern to deform, offering the chance to form contacts. Once the nanorods are bundled together via dopamine, like dopaHexNRA, the nanopillars at the perimeter of the hexagonal pattern could no longer deform, losing the adhesion ability. And, this is also why no adhesion was detected in Hex.

The friction performance against rough surface was also examined [Fig. 5(c)]. Generally, $F_f$ on a rough surface was smaller than that on a smooth surface. $F_f$ of dopaNRA linearly increased from $1.18 \pm 0.03$ to $4.23 \pm 0.05$ mN upon the increase of $F_L$ from 1 to 5 mN, showing a friction coefficient of ~0.76. $F_f$ of dopaNRA under $F_L = 5$ mN is ca. 18% of that on the smooth surface. Both $F_f$ and the friction coefficient are smaller than that on the smooth surface, suggesting the importance of effective contact for the friction. Note that $F_f$ of NRA is larger than dopaNRA under $F_L \leq 3$ mN, while $F_f$ of NRA is smaller under $F_L \geq 3$ mN. Since NRA tends to form close contact with the surface, the result indicates that a better contact contributes mainly to the larger $F_f$ at $F_L \leq 3$ mN. However, for dopaNRA under $F_L \geq 3$ mN, not only is the loading force large enough to endow good contacts but also the incorporation of dopamine imparts a bonding effect to nanorods. Therefore, dopaNRA shows higher friction than NRA at $F_L \geq 3$ mN. Meanwhile, the friction performance of dopaHexNRA, HexNRA, Hex on a rough surface is analogous to a smooth surface. It is worth mentioning that nanorods are easy to bend and collapse under friction test against a rough surface (Fig. 6). Thus, the bonding effect plays a very important role in friction performance.

## IV. SUMMARY AND CONCLUSIONS

Inspired by the tightly connected keratin nanopillars found on the toe pad of tree frog *Staurois parvus*, series of PCL nanorod arrays have been successfully prepared. PCL NRA arrays showed significant enhancement both in adhesion and friction compared to the hexagonal micropillar array (Hex). dopaNRA showed the highest friction that is 2.5 and 2.3 times of the Hex on smooth and rough surfaces, respectively. As for adhesion performance, nanorod arrays and nanorod arrays with hexagon patterns present a relative high value of $0.16 \pm 0.05$ and $0.057 \pm 0.03$ mN on a smooth and rough surface, respectively. The bonding of nanorods with dopamine, which imitates the tightly connected keratin nanopillars on the toe pads of the tree frog, was found to be beneficial to friction rather than to adhesion. The bonding of nanopillars affects the deformability of nanopillars and thus the effective contact area and attaching ability. The results here offer us an opportunity to understand the function of nanostructures on tree frogs' toe pads and to design new bioinspired fibrillar adhesives whether on rough or smooth surfaces.

### ACKNOWLEDGMENTS

This work was supported by the National Natural Science Foundation of China (NNSFC) (No. 51973165) and the National Key R&D Program of China (No. 2018YFB1105100).

### DATA AVAILABILITY

The data that support the findings of this study are available from the corresponding author upon reasonable request.

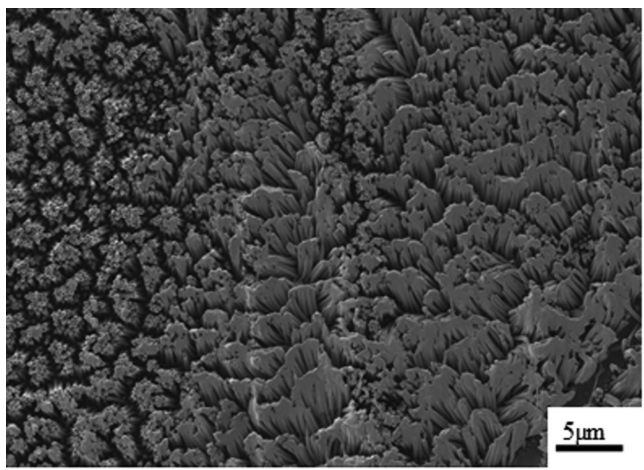

**FIG. 6.** SEM image of dopaNRA after friction measurement with the rough probe.

### REFERENCES

[1]K. Autumn, Y. A. Liang, S. T. Hsieh, W. Zesch, W. P. Chan, T. W. Kenny, R. Fearing, and R. J. Full, Nature **405**, 681 (2000).
[2]V. V. Ernst, Tissue Cell **5**, 97 (1973).
[3]L. Qu, L. Dai, M. Stone, Z. Xia, and Z. L. Wang, Science **322**, 238 (2008).
[4]M. P. Murphy, B. Aksak, and M. Sitti, Small **5**, 170 (2009).
[5]J. K. A. Langowski, D. Dodou, P. Van Assenbergh, and J. L. Van Leeuwen, Integr. Comp. Biol. **60**, 906 (2020).
[6]K. Jin, J. C. Cremaldi, J. S. Erickson, Y. Tian, J. N. Israelachvili, and N. S. Pesika, Adv. Funct. Mater. **24**, 574 (2014).

[7]D. Tan, Y. Zheng, and L. Xue, "The role of effective elastic modulus in the performance of structured adhesives," in *Bio-inspired Structured Adhesives*, edited by L. Heepe, L. Xue, and S. N. Gorb (Springer International Publishing, Cham, 2017), pp. 107–139.

[8]J. K. A. Langowski, D. Dodou, M. Kamperman, and J. L. Van Leeuwen, Front. Zoology **15**, 32 (2018).

[9]X. Wang, D. Tan, X. Zhang, Y. Lei, and L. Xue, Biomimetics **2**, 10 (2017).

[10]J. K. A. Langowski, H. Schipper, A. Blij, F. T. Van Den Berg, S. W. S. Gussekloo, and J. L. Van Leeuwen, J. Anat. **233**, 478 (2018).

[11]Q. Liu *et al.*, Small **17**, 2005493 (2021).

[12]L. F. Boesel, C. Greiner, E. Arzt, and A. del Campo, Adv. Mater. **22**, 2125 (2010).

[13]D. Tan, X. Wang, Q. Liu, K. Shi, B. Yang, S. Liu, Z.-S. Wu, and L. Xue, Small **15**, 1904248 (2019).

[14]X. Wang *et al.*, ACS Appl. Mater. Interfaces **11**, 46337 (2019).

[15]F. Meng, Q. Liu, X. Wang, D. Tan, L. Xue, and W. J. P. Barnes, Philos. Trans. R. Soc. A **377**, 20190131 (2019).

[16]J. M. Smith, W. J. P. Barnes, J. R. Downie, and G. D. Ruxton, J. Comp. Physiol. A **192**, 1193 (2006).

[17]T. Endlein, A. Ji, S. Yuan, I. Hill, H. Wang, W. J. P. Barnes, Z. Dai, and M. Sitti, Proc. R. Soc. B **284**, 20162867 (2017).

[18]W. Federle, W. J. P. Barnes, W. Baumgartner, P. Drechsler, and J. M. Smith, J. R. Soc. Interface **3**, 689 (2006).

[19]D. M. Drotlef, E. Appel, H. Peisker, K. Dening, A. Del Campo, S. N. Gorb, and W. J. Barnes, Interface Focus **5**, 20140036 (2015).

[20]L. Xue *et al.*, ACS Nano **11**, 9711 (2017).

[21]M. Nakano and T. Saino, J. Morphol. **277**, 1509 (2016).

[22]H. Chen, L. Zhang, D. Zhang, P. Zhang, and Z. Han, ACS Appl. Mater. Interfaces **7**, 13987 (2015).

[23]R. Gupta and J. Fréchette, Langmuir **28**, 14703 (2012).

[24]J. Iturri, L. Xue, M. Kappl, L. García-Fernández, W. J. P. Barnes, H.-J. Butt, and A. del Campo, Adv. Funct. Mater. **25**, 1499 (2015).

[25]L. Zhang, H. Chen, Y. Guo, Y. Wang, Y. Jiang, D. Zhang, L. Ma, J. Luo, and L. Jiang, Adv. Sci. **7** 2001125 (2020).

[26]Q. Liu *et al.*, ACS Appl. Mater. Interfaces **12**, 19116 (2020).

[27]A. Tsipenyuk and M. Varenberg, J. R. Soc. Interface **11**, 20140113 (2014).

[28]H. Ko, M. Seong, and H. E. Jeong, Soft Matter **13**, 8419 (2017).

[29]C. Dhong and J. Fréchette, Soft Matter **11**, 1901 (2015).

[30]L. Pang, N. C. Paxton, J. Ren, F. Liu, H. Zhan, M. A. Woodruff, A. Bo, and Y. Gu, ACS Appl. Mater. Interfaces **12**, 47993 (2020).

[31]H. Masuda, K. Yada, and A. Osaka, Jpn. J. Appl. Phys. **37**, L1340 (1998).

[32]W. Cheng, X. Zeng, H. Chen, Z. Li, W. Zeng, L. Mei, and Y. Zhao, ACS Nano **13**, 8537 (2019).

[33]L. Xue, J. Iturri, M. Kappl, H.-J. Butt, and A. del Campo, Langmuir **30**, 11175 (2014).

[34]D.-M. Drotlef, L. Stepien, M. Kappl, W. J. P. Barnes, H.-J. Butt, and A. del Campo, Adv. Funct. Mater. **23**, 1137 (2013).

[35]C. Zhang, Y. Ou, W.-X. Lei, L.-S. Wan, J. Ji, and Z.-K. Xu, Angew. Chem. Int. Ed. **55**, 3054 (2016).

[36]L. Xue, A. Kovalev, F. Thöle, G. T. Rengarajan, M. Steinhart, and S. N. Gorb, Langmuir **28**, 10781 (2012).

[37]L. Xue, A. Kovalev, K. Dening, A. Eichler-Volf, H. Eickmeier, M. Haase, D. Enke, M. Steinhart, and S. N. Gorb, Nano Lett. **13**, 5541 (2013).